# An Improved Dipole Extraction Method From Magnitude-Only Electromagnetic-Field Data

Chunyu Wu, Ze Sun, Xu Wang, Yansheng Wang, Ben Kim and Jun Fan, *Fellow, IEEE*

*Abstract*—**Infinitesimal electric and magnetic dipoles are widely used as an equivalent radiating source model. In this paper, an improved method for dipole extraction from magnitude-only electromagnetic-field data based on genetic algorithm and back-and-forth iteration algorithm [1] is proposed. Compared with conventional back-and-forth iteration algorithm, this method offers an automatic flow to extract the equivalent dipoles without prior decision of the type, position, orientation and number of dipoles. It can be easily applied to electromagnetic-field data on arbitrarily shaped surfaces and minimize the number of required dipoles. The extracted dipoles can be close to original radiating structure, thus being physical. Compared with conventional genetic algorithm based method, this method reduces the optimization time and will not easily get trapped into local minima during optimization, thus being more robust. This method is validated by both simulation data and measurement data and its advantages are proved. The potential application of this method in phase retrieval is also discussed.**

## I. INTRODUCTION

INFINITESIMAL dipoles are widely used as an equivalent model of radiating source in far field radiated emissions, near-field coupling and antenna modelling [2] areas. The infinitesimal dipoles which generate the same radiated electromagnetic field can be used to replace the original radiating noise source.

These equivalent dipoles can be directly applied to calculation of the far field value as shown in [2][3] since radiation from infinitesimal dipoles is well known. This implies the application of equivalent dipoles in NF to FF transformation. They can also be imported in full wave simulation tool to simulate the far field value of noise source with added enclosure [4] or near-field coupling from noise source to a victim antenna [5]. In [6], a physics-based dipole is extracted and used to debug the radiation mechanism of a flexible printed circuit board (PCB). It's found that physical dipoles which are close to original radiation source are useful in terms of finding the noise current path. Magnetic dipoles are formed by current loops which may incorporate displacement current. However, the physical dipoles are extracted by recognizing the pattern generated by a single horizontal magnetic dipole. In other cases, the pattern may be complicated and cannot be recognized easily. Later, a dipole based reciprocity method is proposed to calculate the coupled noise from a physical dipole to victim antenna [7]. The direction and location of the physical dipole can be optimized to reduce the coupled noise. This can provide guideline to optimize the placement of noise source and layout in actual DUT. In conclusion, physical dipoles turn out to be useful in debugging of radiation mechanism (current flow), near-field coupling estimation and its reduction.

There are several methods to extract equivalent dipoles from scanned radiated electromagnetic field. In [3], genetic algorithm is used to optimize the type (electric or magnetic), position, orientation, magnitude and phase of all dipoles. This method works pretty well but the disadvantages are that it is very time consuming because of the large number of optimization variables (8 variables for each dipole) and the optimization can fall into local minima easily because of the large variation range of dipole magnitude which is from zero to positive infinity. As a result, the number of dipoles is usually strictly limited. In [8], a set of vertical magnetic dipoles or horizontal electric dipoles are placed on the discretized horizontal surface of a radiating integrated circuit (IC) and linear least square method is used to solve the current intensity of these dipoles. The method requires simple matrix operations to obtain the results and is very fast. But the dipoles obtained do not correspond to the real source and the position and number of dipoles need to be determined in advance by meshing the surface of IC. In [9], an array of uniformly placed dipole sets are used and each dipole set includes one vertical electric dipole $P^z$ and two horizontal magnetic dipoles $M^x$, $M^y$ at the same position. The regularization technique and the truncated singular-value decomposition method are investigated together with the conventional least-squares method to calculate the dipole magnitude and phase from the near-field data. The regularization technique can achieve a better source model that reflects the actual physics. But the physics is shown by the 2-D pattern of dipole magnitude and phase, so a large number of

This paragraph of the first footnote will contain the date on which you submitted your paper for review. It will also contain support information, including sponsor and financial support acknowledgment. For example, "This work was supported in part by the U.S. Department of Commerce under Grant BS123456."

The next few paragraphs should contain the authors' current affiliations, including current address and e-mail. For example, F. A. Author is with the National Institute of Standards and Technology, Boulder, CO 80305 USA (e-mail: author@ boulder.nist.gov).

S. B. Author, Jr., was with Rice University, Houston, TX 77005 USA. He is now with the Department of Physics, Colorado State University, Fort Collins, CO 80523 USA (e-mail: author@lamar.colostate.edu).

T. C. Author is with the Electrical Engineering Department, University of Colorado, Boulder, CO 80309 USA, on leave from the National Research Institute for Metals, Tsukuba, Japan (e-mail: author@nrim.go.jp).

dipole sets is needed. Also, how to select the regularization coefficient brings another problem. By combining the genetic algorithm and linear least square method, the equivalent dipoles are obtained with a reduced number of dipoles in a reduced computing time without prior decision of the location and number of dipoles in [10], but the algorithm does not optimize the dipole type. It was assumed that only magnetic dipoles exist.

Since the methods involving matrix inversion require both magnitude and phase of electric or magnetic field while phase-resolved electric or magnetic field measurement brings more difficulty than magnitude-only electric or magnetic field measurement, Ji proposes a back-and-forth iteration algorithm which requires only the information of the field magnitudes on two near-field scanning planes to extract dipoles [1]. This method is effective and fast to extract dipoles from magnitude-only near-field data, but there are two disadvantages about this method. The first disadvantage is that the method requires propagating field from the lower plane to the higher plane. This will bring some difficulties when cylindrical scan is used. Transformation between field on two different cylindrical surfaces is much more difficult than transformation between field on two different planar surfaces. And it requires that the sampling must be done uniformly on the surface. The second disadvantage is that the dipole sets which include one vertical electric dipole $P^z$ and two horizontal magnetic dipoles $M^x$, $M^y$ at the same position are placed uniformly. This will lead to a large number of dipoles and more scanning points to have a well-conditioned transformation matrix during transformation from field to dipole moments. Also, the position and number of dipole sets need to be determined in advance. This is usually done by trial and error.

In this paper, an improved method for dipole extraction from magnitude-only electromagnetic-field data based on genetic algorithm and back-and-forth iteration algorithm is proposed, which aims at solving the two disadvantages mentioned above. It offers an automatic flow to extract the equivalent dipoles without prior decision of the type, position, orientation and number of dipoles. Compared with conventional back-and-forth iteration algorithm, this method can be easily applied to electromagnetic-field data on arbitrarily shaped surfaces and minimize the number of dipoles. It can also generate physical dipoles which are close to original radiating source. Compared with conventional genetic algorithm based method, this method reduces the optimization time and will not easily get trapped into local minima during optimization because it will not optimize the magnitude and phase of dipoles.

This paper is organized as below. Section II introduces the method in detail. Section III validates the method using simulated magnitude-only electromagnetic-field data on cylindrical surfaces with infinitesimal dipoles or wire antenna as source and measured magnitude-only electromagnetic-field data on cylindrical surfaces with a radiating server as source. Section IV extends the method into single surface scanning and discusses its potential application in phase retrieval. Section V concludes the paper.

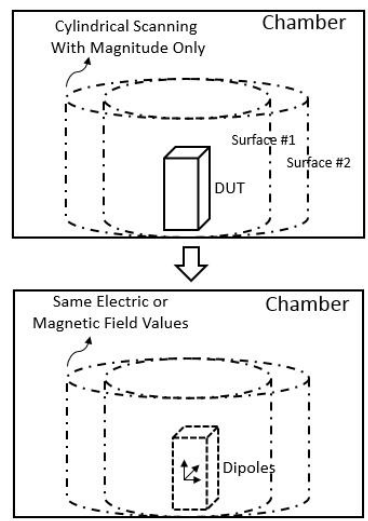

Fig. 1. Principle of the method.

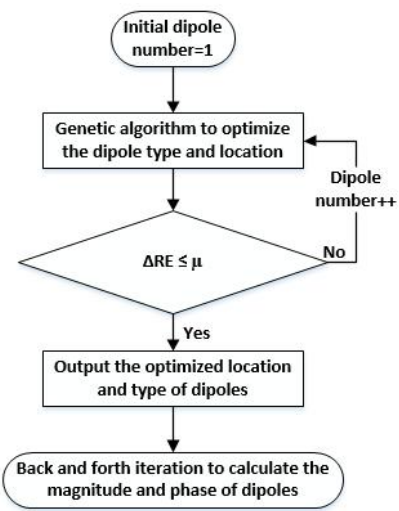

Fig. 2. General flow of the method.

## II. DESCRIPTION OF THE METHOD

The principle of this method is shown in Fig. 1. Electric or magnetic field data with magnitude only are obtained firstly on two cylindrical surfaces around DUT, then infinitesimal dipoles are extracted to generate the same electric or magnetic field. These infinitesimal dipoles can be considered as an equivalent radiating source model. Although the method is discussed and validated in term of cylindrical scanning in this paper, it can be applied easily to planar scanning and spherical scanning as well. Electric field is taken as input data to describe and validate the method in this paper, but the input data can also be magnetic field.

The general flow of this method is shown in Fig. 2. The method starts with one dipole. Then genetic algorithm is used to optimize the dipole type and dipole position to minimize the relative error between the measured electric field and calculated electric field from equivalent dipole source. After optimization, the minimized relative error will be compared with previous relative error with one less dipole. If the decrease of relative error is smaller or equal than μ, it can be concluded that the relative error has converged and it will output the optimized location and type of dipoles. Otherwise, the number of dipoles will be increased by one and this optimization will be executed again. μ means the minimum decrease of relative error which can be tolerated with increment of dipole number. The definition of relative error is shown in equation (1)~(3), where RE1 is the relative error between the measured electric field and calculated electric field from equivalent dipole source in surface #1, RE2 is the relative error in surface #2 and RE is the overall relative error which is the average of RE1 and RE2.

$$RE1 = \sqrt{\frac{\sum_{Surface\ \#1}[(\left|E_z^{scan}\right| - \left|E_z^{fit}\right|)^2 + (\left|E_\varphi^{scan}\right| - \left|E_\varphi^{fit}\right|)^2]}{\sum_{Surface\ \#1}(\left|E_z^{scan}\right|^2 + \left|E_\varphi^{scan}\right|^2)}} \quad (1)$$

$$RE2 = \sqrt{\frac{\sum_{Surface\ \#2}[(\left|E_z^{scan}\right| - \left|E_z^{fit}\right|)^2 + (\left|E_\varphi^{scan}\right| - \left|E_\varphi^{fit}\right|)^2]}{\sum_{Surface\ \#2}(\left|E_z^{scan}\right|^2 + \left|E_\varphi^{scan}\right|^2)}} \quad (2)$$

$$RE = \frac{1}{2}(RE1 + RE2) \quad (3)$$

### A. Improved Back-and-forth Iteration Algorithm

The improved back-and-forth iteration algorithm is shown in Fig. 3. The algorithm starts with initial values of dipole magnitude and phase and given dipole position and type. The initial values can be decided by assuming that the field at every scanning point on surface #2 has the same phase. Then the initial values are obtained using linear least square method.

*step 1*: calculate the field on surface #1 using the transfer function relating field on surface #1 with dipole source as shown in equation (4), where F represents the field value vector, D represents the dipole value vector and T represents the transfer function which is determined by the location and types of dipoles and the location of scanning points.

$$[F]_{M \times 1} = [T]_{M \times N}[D]_{N \times 1} \quad (4)$$

*step 2*: calculate the relative error on surface #1, namely RE1.

*step 3*: enforce the magnitude of field on surface #1 to be the measured magnitude but keep the phase unchanged.

*step 4*: use the updated field on surface #1 to inversely calculate the dipole magnitude and phase by linear least square method as shown in equation (5).

$$D = (T^T T)^{-1} T^T F \quad (5)$$

*step 5*: calculate the field on surface #2 using the transfer function relating field on surface #2 with dipole source similar to equation (4).

*step 6*: calculate the relative error on surface #2, namely RE2.

*step 7*: enforce the magnitude of field on surface #2 to be the measured magnitude but keep the phase unchanged.

*step 8*: use the updated field on surface #2 to inversely calculate the dipole magnitude and phase by linear least square method similar to equation (5).

The overall relative error RE will be compared with RE in previous iteration. If the decrease of RE is smaller or equal than ε, it can be concluded that RE has converged and it will output the obtained magnitude and phase of dipoles. Otherwise, this iteration will be executed again. ε means the minimum decrease of RE which can be tolerated with increment of iteration number.

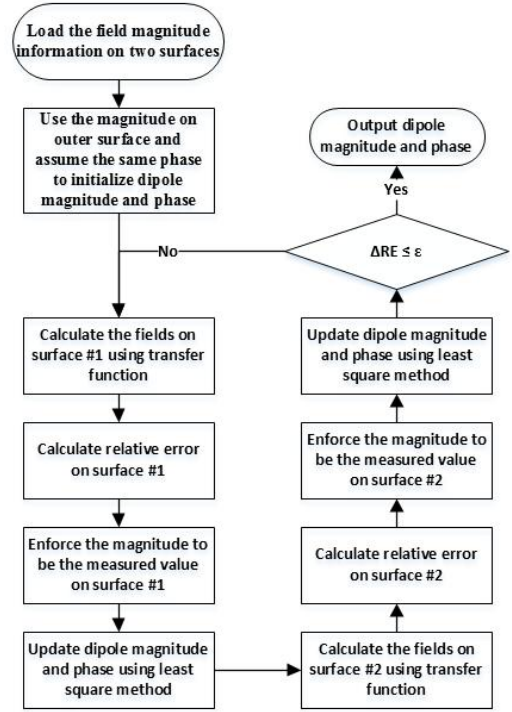

Fig. 3. General flow of the improved back-and-forth iteration algorithm.

The main difference between the improved back-and-forth iteration algorithm introduced in this paper and the conventional algorithm is that transformation between field on two different surfaces is not needed. This will bring two advantages. The first advantage is that the method can be applied to electromagnetic-field scanning in arbitrarily shaped surfaces. In [1], the field transformation from the lower plane to the higher plane is achieved using the plane-wave expansion method. For cylindrical scanning, cylindrical wave expansion which is more complicated is needed. The improved back-and-forth iteration algorithm does not require complicated wave expansion method and can be easily applied in planar, cylindrical or spherical scanning. It is even possible that one scanning is done in planar surface and the other scanning is done in cylindrical surface. The second advantage is that there are no strict requirements about sampling in each surface for the improved algorithm. Since the wave expansion method requires FFT or IFFT to efficiently perform the numerical evaluation of the integral equations converting EM field between spectral domain and spatial domain, the conventional algorithm enforces some sampling restrictions.

### B. Genetic Algorithm

The genetic algorithm is a method for solving both constrained and unconstrained optimization problems that is based on natural selection, the process that drives biological evolution. The genetic algorithm repeatedly modifies a population of individual solutions. At each step, the genetic algorithm selects individuals from the current population to be parents and uses them to produce the children for the next generation. Over successive generations, the population "evolves" toward an optimal solution [11]. It's useful in terms of optimization for highly nonlinear problems. Genetic algorithm is used to optimize the dipole type and location in this method.

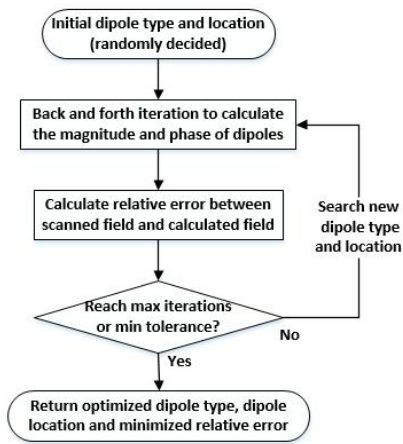

Fig. 4. General flow of the genetic algorithm.

TABLE I
COMPARISON BETWEEN ORIGINAL SOURCE AND EXTRACTION RESULTS

| | | ORIGINAL SOURCE | | EXTRACTED DIPOLES | |
|---|---|---|---|---|---|
| | | DIPOLE 1 | DIPOLE 2 | DIPOLE 1 | DIPOLE 2 |
| DIPOLE LOCATION | X | 0.25 | -0.25 | 0.25 | -0.25 |
| | Y | 0 | 0 | -1.246E-5 | -1.449E-5 |
| | Z | 1.5 | 1.5 | 1.4999 | 1.4998 |
| DIPOLE TYPE | | $P_X$ | $M_Y$ | $P_X$ | $M_Y$ |
| DIPOLE AMPLITUDE | | 1 AM | 100 VM | 0.9999 AM | 99.9968 VM |
| DIPOLE PHASE (DEGREE) | | 90 | 0 | 147.0478 | 57.0494 |

The objective function to minimize is the relative error defined in equation (3). The optimization range for dipole location can be set as the occupied space of original DUT or radiating structure. The optimization range for dipole type can be set as $[P^x, P^y, P^z, M^x, M^y, M^z]$. It means that the dipole type can be any one of the six kinds. Meaningful optimization range can help us extract dipoles close to original radiating source and reduce optimization time. For example, the radiation from horn antenna is equivalent as radiation from $\mathbf{J}$ and $\mathbf{M}$ on the aperture, then the optimization range of dipole location can also be set as the aperture surface of horn antenna.

The general flow of the optimization using genetic algorithm is shown in Fig. 4. The algorithm starts with a population of randomly decided individuals for dipole type and location. Then the improved back-and-forth iteration algorithm is used to obtain the magnitude and phase of dipoles for each individual. The objective function for each individual is evaluated next for selection of parent. Subsequent generations evolve from the current generation through selection, crossover and mutation to search new dipole type and location. And this procedure will go over and over again until max number of generations is reached or the minimized value of objective function is no longer changing. Then the algorithm will stop and return the optimized dipole type, dipole location and minimized relative error.

## III. VALIDATION OF THE METHOD

### A. Simulation Data with Infinitesimal Dipoles as Source

The method is firstly validated using simulation data with infinitesimal dipoles as source. In this case, two infinitesimal dipoles are used as the original radiating source. It is expected that the proposed algorithm can extract the same infinitesimal dipoles. One Px and one My dipole are placed above the PEC ground. The simulated E field (Ez and Ephi) magnitude on two cylindrical surfaces with radius of 0.5m and 1m at 781.25MHz is used as the input data to the algorithm. The scanning height is from 1m to 4m with step of 0.25m and the scanning points along circumference is shown in Fig. 5. The optimization range for x coordinate of dipole location is from -0.5 to 0.5, The optimization range for y coordinate of dipole location is from -0.5 to 0.5 And The optimization range for z coordinate of dipole location is from 1 to 2. The extraction results are shown in Table I.

The relative error between the calculated fields from extracted dipoles and scanned fields is 0.00035. The extracted dipoles are almost the same as original radiating source in terms of dipole number, dipole type, dipole location and dipole amplitude. The phase difference of extracted two dipoles are 90 degrees, which is the same as original dipole source. The relative error change during back-and-forth iteration to calculate magnitude and phase of dipoles is shown in Fig. 6. The iteration process converges very fast.

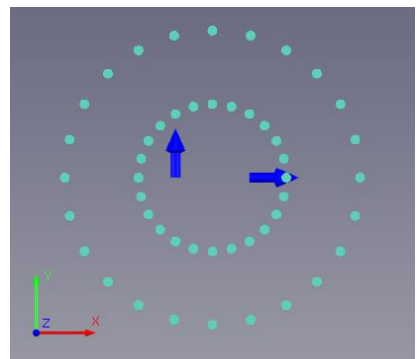

Fig. 5. Scanning points along circumference

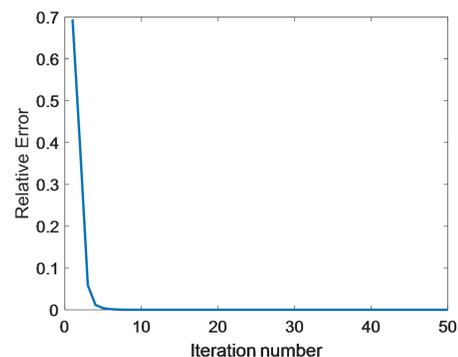

Fig. 6. Relative error change with iteration number

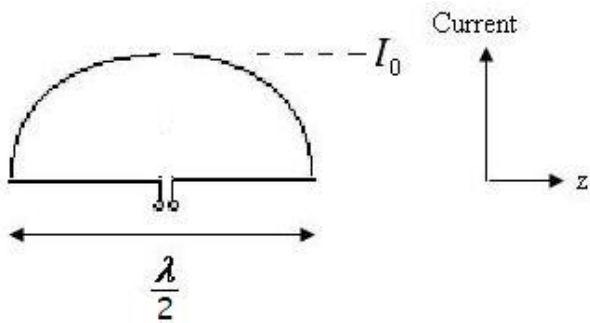

Fig. 7.  Current along half wavelength wire antenna.

### B. Simulation Data with Wire Antenna as Source

To further validate that this algorithm can extract physical dipoles, a half wavelength wire antenna is used as original radiating source. The center of wire antenna is located at (0, 0, 1.5m) and the wire antenna is placed along z axis. The excitation voltage is 5V, 0 degree. The simulated E field (Ez and Ephi) magnitude on two cylindrical surfaces with radius of 0.5m and 1m at 781.25MHz is used as the input to the algorithm. The scanning points distribution is the same as III.A. The optimization range of dipole location is also the same as III.A. The extracted dipole is a single Pz dipole located at (-3.629e-05, 2.226e-06, 1.4977) with magnitude of 0.0066 and phase of -10.17 degrees. The relative error between the calculated fields from dipole and scanned fields is 0.0339. The extracted dipole is close to original radiating structure since the current in original wire antenna flows in the z direction and the radiating wire antenna is composed of a series of Pz dipoles along the wire. Because the observation plane is far away from wire antenna, a series of Pz dipoles along the wire can be approximately equivalent to a single Pz dipole in the center.

The current along the half wavelength wire antenna can be expressed as equation (4), as shown in Fig. 7.

$$I(z) = I_0 \sin\left[\frac{2\pi}{\lambda}\left(\frac{\lambda}{4} - z\right)\right], \quad -\frac{\lambda}{4} \le z \le \frac{\lambda}{4} \quad (4)$$

$$\text{where } I_0 = \frac{V}{Z_{in}} = \frac{5 \, V}{(73 + j42.5) \, Ohm}$$

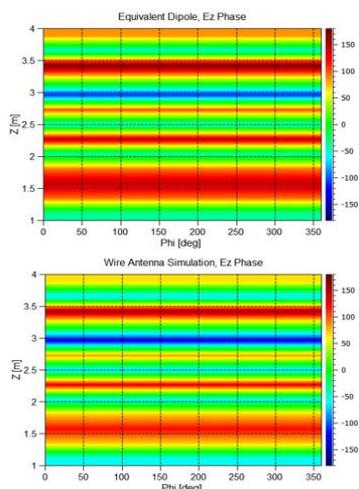

Fig. 8.  Comparison of retrieved phase pattern and original simulated phase pattern on cylindrical surface.

The value of Pz dipole can be calculated using equation (5). The calculated value is very close to the value of extracted single Pz dipole.

$$\int_{-\frac{\lambda}{4}}^{\frac{\lambda}{4}} I(z)dz = \frac{I_0 \lambda}{\pi} = \frac{V\lambda}{Z_{in}\pi} = 0.0072 \quad (5)$$

The phase pattern of calculated Ez field on the cylindrical surface with radius of 0.5m is compared with original wire antenna simulation as shown in Fig. 8. The phase pattern is retrieve correctly.

### C. Measurement Data with Radiating Rack as Source

To further validate the algorithm, a data center rack is used as DUT. The horizontal (Ephi) and vertical (Ez) E field generated by the rack on two cylindrical surfaces with radius of 2m and 5m is measured in a semi-anechoic chamber. The measurement setup is shown in Fig. 9.

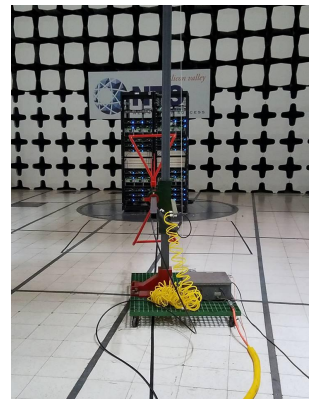

Fig. 9.  Electric field scan of a rack.

The measured E field pattern at 781.25MHz is shown in Fig. 10. The unit is dBuV/m. The x axis is the azimuth in degree and the y axis is the height.

The measured E field magnitude is used as input to the algorithm. The algorithm extracted 9 dipoles with a relative error of 0.1885. The calculated E field pattern on the same cylindrical surfaces from extracted dipoles is shown in Fig. 11.

To compare the measured E field and calculated E field from extracted dipoles, the first column in Fig. 9 and Fig. 10 (Ez and Ephi, r=2m, phi=0 degree) is plotted in one figure as shown in Fig. 12.

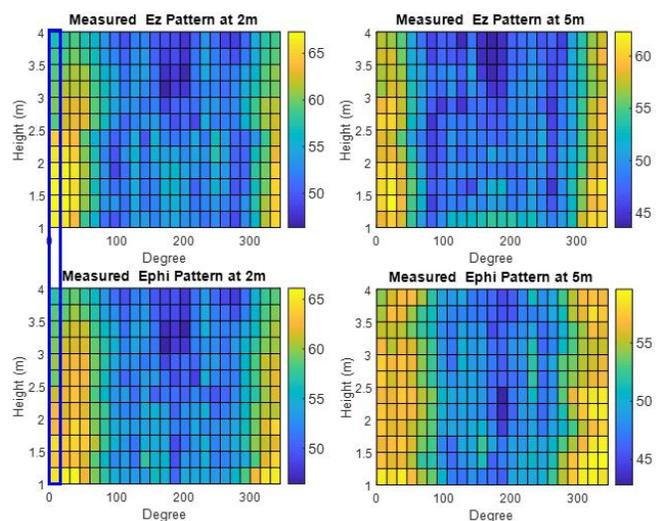

Fig. 10.  Scanned electric field pattern of a rack.

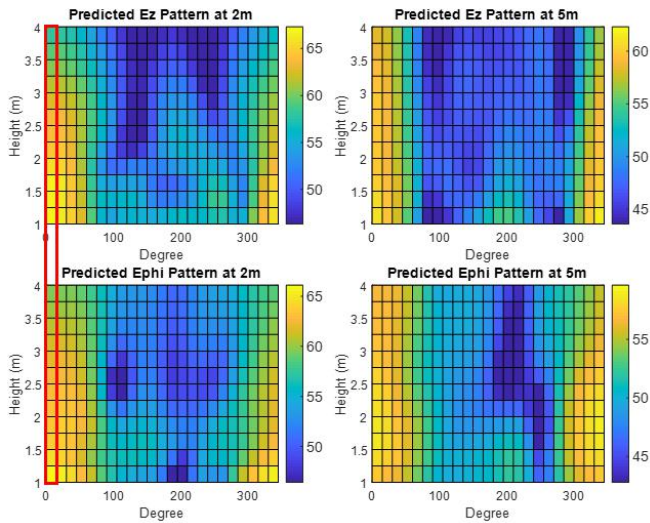

Fig. 11. Predicted electric field pattern from extracted dipoles.

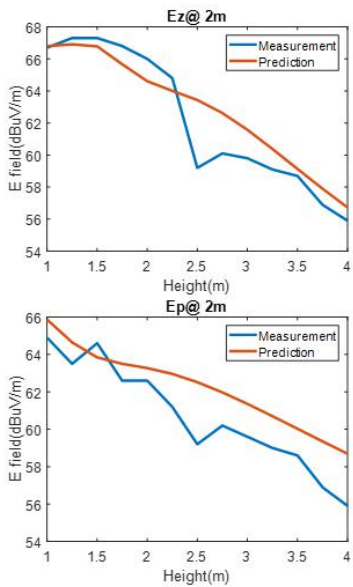

Fig. 12. Comparison of measured E field and predicted E field.

From the comparison, we can see that the calculated E field from extracted dipoles matches well with measurement. The algorithm can tolerate a certain amount of noise in original measurement and will not over fit the noise. The predicted E field pattern is smoother with less noise.

## IV. EXTENSIONS

### A. Single Surface Scanning

Although in the back-and-forth iteration algorithm, field magnitude on two surfaces is used to get the dipole value, the algorithm can still work if field on only one surface is available. The algorithm can be adapted as shown in Fig. 13 if the field magnitude on only one surface is available. The reason why field magnitude on two surfaces are normally used is to provide more information about the radiating source for extracting dipoles close to original radiating source and reduce the influence of possible measurement error. But if measurement data on the second surface does not provide more information or it takes too much time to scan on the second surface, scanning data on only one surface can also be used with the algorithm.

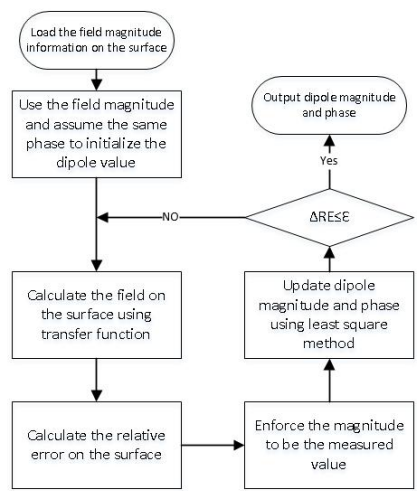

Fig. 13. General flow of the back-and-forth iteration algorithm for single-surface scanning.

The example in III.B is used again to validate the algorithm with single scanning surface. Ez and Ephi on cylindrical surface with radius of 1m is used as input to the algorithm. The extracted dipole is still a Pz dipole at (5.4561e-8, 1.9085e-7, 1.494) with magnitude of 0.0066 and phase of -10.3007 degrees. The reason why single surface scanning works well for this case is the source is very simple and electric field data on only one surface already provides enough information about the source.

As shown in Fig. 14, the uniqueness theorem [12] stated that if the tangential components of E or H field over the boundary S and the sources J and M are specified, field in the region is unique when the frequency does not equal the characteristic frequency. If the frequency equals the characteristic frequency, the derivative of the tangential components of the field over the boundary with respect to frequency should also be given in order to obtain the unique field.

In the source reconstruction case with magnitude-only electromagnetic-field data, obviously the uniqueness theorem does not hold anymore. This means that there might exist several solutions of equivalent dipoles inside the scanning surface which can generate the same scanning field magnitude. In order to generate equivalent dipoles which are close to original radiating structure, more information about the source needs to be provided. This is the reason why two or three scanning surfaces and specification of the range of dipole location or type will help extract dipoles close to original radiating structure.

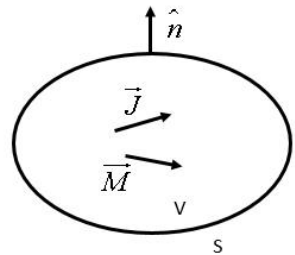

Fig. 14. Arbitrarily shaped surface S encloses linear lossless matter and sources.

### B. Potential Application in Phase Retrieval

There are many papers studying phase retrieval from magnitude-only electromagnetic-field data [13]-[16]. These methods work well but they are all limited on planar near-field measurement and require two planes or two probes during measurement. They also impose restrictions on the sampling points on the plane. The proposed method in this paper provides another way to retrive phase from magnitude-only electromagnetic-field measurement. When the equivalent dipoles are extracted, the phase on the measurement surfaces can also be easily calculated from extracted dipoles. Since this method can be applied to electromagnetic-field data on arbitrarily shaped surfaces, this method can retrieve phase for phaseless measurement not only on planar surfaces, but also on cylindrical surfaces or spherical surfaces. The sampling points can be sparse, not nessarily be uniform and follow a certain pattern.

## V. Conclusion

This paper proposes an improved method to extract dipoles from magnitude-only electromagnetic-Field data. It offers an automatic flow to extract the equivalent dipoles without prior decision of the type, position, orientation and number of dipoles. It can be easily applied to electromagnetic-field data on arbitrarily shaped surfaces and minimize the number of required dipoles. The extracted dipoles can be close to original radiating structure, thus being physical if adequate field magnitude information is acquired and meaningful optimization range of dipole location and type is given. Compared with conventional genetic algorithm based method which optimize every parameter of dipoles, this method reduces the optimization time and will not easily get trapped into local minima during optimization, thus being more robust. It also shows good potential in phase retrieval application.

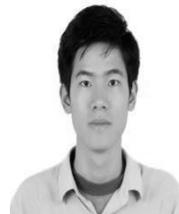

**Chunyu Wu** received his Bachelor's degree in Electrical Engineering from Huazhong University of Science and Technology, Wuhan, China in 2014. He received his Master's degree in Electrical Engineering from Missouri University of Science and Technology, MO, USA in 2017. He is currently working toward his Ph.D. degree in Electrical Engineering at the EMC Laboratory, Missouri University of Science and Technology.

His research interest includes interference control, signal integrity and power integrity.

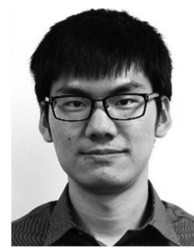

**Yansheng Wang** (S'11) received the B.S. degree in electrical engineering from Beihang University, Beijing, China, in 2012 and Ph.D. degree in electrical engineering at EMC Laboratory, Missouri University of Science and Technology, Rolla, MO, USA in 2018.

His research interests include radiation characterization for cable harness inside vehicle body, method of moments, multiple-input-multiple-output over-the-air testing, desense measurement and simulation, radio-frequency interference, broadband Hfield probe design, conducted emission modeling for switched-mode power supply, stochastic modeling for high-speed channels, and generic model development for PCIe3/4.

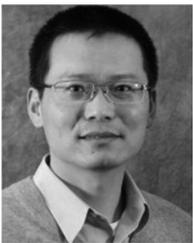

**Jun Fan** (S'97–M'00–SM'06–F'16) received the B.S. and M.S. degrees in electrical engineering from Tsinghua University, Beijing, China, in 1994 and 1997, respectively, and the Ph.D. degree in electrical engineering from the University of Missouri-Rolla, Rolla, MO, USA, in 2000.

From 2000 to 2007, he was a Consultant Engineer with NCR Corporation, San Diego, CA, USA. In July 2007, he joined Missouri University of Science and Technology (formerly University of Missouri-Rolla), where he is currently a Professor and Director of Missouri Science and Technology Electromagnetic Compatibility (S&T EMC) Laboratory. He is also the Director of the National Science Foundation Industry/

University Cooperative Research Center for EMC and Senior Investigator of Missouri S&T Material Research Center. His research interests include signal integrity and EMI designs in high-speed digital systems, dc power-bus modeling, intrasystem EMI and RF interference, printed circuit board noise reduction, differential signaling, and cable/connector designs.

Dr. Fan was the Chair of the IEEE EMC Society TC-9 Computational Electromagnetics Committee from 2006 to 2008 and a Distinguished Lecturer of the IEEE EMC Society in 2007 and 2008. He is currently the Chair of the Technical Advisory Committee of the IEEE EMC Society and is an Associate Editor of the IEEE TRANSACTIONS ON ELECTROMAGNETIC COMPATIBILITY and EMC Magazine. He received an IEEE EMC Society Technical Achievement Award in August 2009.